- **Submitted to The Journal of Physical Chemistry C**

# First-Principles Study of the Temperature Dependence of Structural, Electronic, and Hyperfine Properties of the Cu(100) Surface

*Germán N. Darriba*[*,†], *Ricardo Faccio*[§] *and Mario Rentería*[*,†]

† *Departamento de Física "Prof. Dr. Emil Bose" and Instituto de Física La Plata [IFLP, Consejo Nacional de Investigaciones Científicas y Técnicas (CONICET) La Plata], Facultad de Ciencias Exactas, Universidad Nacional de La Plata, CC 67, (1900) La Plata, Argentina.*

§ *Área Física y Centro NanoMat, DETEMA, Facultad de Química, Universidad de la República, Gral. Flores 2124, P.O. Box 1157, Montevideo, Uruguay.*

KEYWORDS: Electronic structure, TDPAC, DFT calculations, EFG, Surface.

ABSTRACT: In this work, we investigate the temperature-dependent behavior of the pure (undoped) Cu(100) surface using first-principles calculations within the Density Functional Theory framework. One of the main objectives is to determine, as a first step, whether the unexpected linear dependence of the electric-field gradient (EFG) tensor observed in TDPAC experiments on $^{111}$Cd probes deposited on Cu(100) surfaces could arise from the surface generation itself or from surface reconstruction. To this end, we perform a comprehensive *ab initio* study of the Cu(100) surface reconstruction and its associated structural, electronic, and hyperfine properties as a function of temperature, not only at the outermost atomic layer (i.e., the topmost Cu atom) but also as a function of atomic depth relative to the surface. To study the temperature dependence of the EFG, we use experimentally determined temperature-dependent lattice parameters for bulk copper in our calculations. The anisotropic relaxation that arises when bulk symmetry is broken helps unravel the potential sources of EFG temperature dependence at the surface. Studying the electron density of conduction electrons $\rho(\vec{r})$ at the atomic scale near the Cu nucleus and the atom-resolved partial density of states at the topmost Cu atom allows us to correlate the surface effect on the EFG with respect to the bulk value. Finally, we discuss the linear temperature dependence found for the EFG on the undoped Cu(100) surface in relation to the temperature dependence of the "lattice" contribution to the EFG in the framework of the universal correlation found by Raghavan *et al*. [Phys. Rev. Lett. 1975, 34, 1280] for noncubic metals. Finally, we show that the linear decrease of the EFG as T increases in the $^{111}$Cd-doped Cu(100) surface experiments is probably mostly caused by the generation of the pure Cu(100) surface and its subsequent reconstruction.

## I. INTRODUCTION

The atomic-scale geometric configuration at and near solid surfaces plays a pivotal role in understanding interatomic and interlayer interactions under these boundary conditions. When surfaces are generated, several new physical properties that are not exhibited in bulk materials can emerge, significantly influencing their behavior and potential technological applications [1]-[5].

Surface formation inherently breaks the bulk crystal's translational symmetry, generating stress fields that drive structural relaxation and reconstruction toward new equilibrium configurations. These *reconstructed* surfaces often exhibit distinct electronic and magnetic properties compared with their bulk counterparts, making their characterization essential for both fundamental and applied research.

Hyperfine interactions with suitable radioactive probe atoms are a powerful tool for studying atomic environments and electronic and magnetic properties in solids. Magnetic dipole and electric quadrupole hyperfine couplings provide insight into exchange interactions and the spatial distribution of electron charge density surrounding atomic probe nuclei. In particular, the electric-field gradient (EFG) tensor at the site of the probe nucleus is highly sensitive to the anisotropy of the local electronic environment due to its inverse-cubic dependence on the radial coordinate, $r^{-3}$. The EFG components are defined as $V_{ij}(\vec{r}) = \frac{\partial^2 V(\vec{r})}{\partial x_i \partial x_j}$, where $V(\vec{r})$ is the electrostatic potential. In its principal axis system, the diagonalized EFG tensor (being traceless) is fully characterized by its largest principal component $V_{33}$ and the asymmetry parameter $\eta = \frac{V_{11} - V_{22}}{V_{33}}$, with the convention standard $|V_{11}| \leq |V_{22}| \leq |V_{33}|$, reflecting the local electron density anisotropy. The highly precise experimental measurement of this tensor, combined with reliable EFG calculations, can offer comprehensive insight into the electron density (represented as $\rho(\vec{r})$) of the system being investigated or designed. The EFG at a probe nucleus is determined with high precision using time-differential perturbed $\gamma$-$\gamma$ angular correlations (TDPAC) spectroscopy. This technique measures the nuclear quadrupole frequency $\omega_Q$ of the hyperfine interaction between the EFG tensor and the nuclear quadrupole moment $Q$ of the probe´s sensitive nuclear state. Then, $V_{33}$ is determined through $\omega_Q = eQV_{33}/40\hbar$ (for the most widely used $^{111}$Cd probe studied here), provided the quadrupole moment $Q$ is known. Comprehensive descriptions of TDPAC principles and experimental measurements can be found in Refs. [6]-[9].

On FCC metal surfaces, numerous experimental [10]-[17] and first-principles [18] studies have investigated the EFG at substitutional sites of diluted $^{111}$Cd probe atoms located at the topmost atomic positions, a measurement that is experimentally challenging. However, only a few of these surfaces have been studied experimentally as a function of measurement temperature. In particular, the study of deposited $^{111}$In($\rightarrow^{111}$Cd) impurities on the Cu(100) surface, using TDPAC spectroscopy, reveals an unexpected and strong linear temperature dependence of the EFG [10]. The relative decrease of $V_{33}$ shown across an extrapolated temperature range from 0 to 1000 K gives 10.5 %. Due to the cubic symmetry of the Cu FCC lattice, the EFG in the bulk is zero, making a direct comparison (in % of change) between bulk and surface behavior impossible. On the other hand, for many bulk materials with noncubic metallic structures (thus with nonnull EFGs), the bulk EFG has a $-\alpha T^{3/2}$dependence (α > 0) [19], with the mean-square displacement of probe atoms and their immediate neighbors responsible for this effect.

In this work, we investigate the temperature-dependent behavior of the pure (undoped) Cu(100) surface using first-principles calculations within the Density Functional Theory (DFT) framework. One of the main goals is to elucidate whether the previously observed linear dependence of the EFG at the $^{111}$Cd-doped Cu(100) surface [1] arises intrinsically during surface formation and reconstruction, is induced by the Cd impurity´s electronic configuration, its deformation, and the local structural relaxation it produces, or is the result of all these effects. To this end, and as a first step toward a more complex future study of the doped surface, we perform here a comprehensive *ab initio* study of the Cu(100) surface reconstruction and its associated structural, electronic, and hyperfine properties as a function of temperature, not only at the outermost atomic layer (i.e., the topmost Cu atom) but also as a function of atomic depth relative to the reconstructed surface.

Understanding surface reconstruction and studying the EFG behavior at all Cu sites as a function of temperature, using experimentally determined temperature-dependent lattice parameters for bulk copper in our calculations and accounting for anisotropic relaxation that arises when bulk symmetry is broken, helps unravel the potential sources of EFG temperature dependence at the doped surface. Additionally, studying the electron density of conduction electrons at the atomic scale near the Cu nucleus and the atom-resolved partial electronic density of states (PDOS) at the topmost Cu atom allows us to correlate the surface effect on the EFG with respect to the bulk value. Finally, we discuss the linear temperature dependence found for the EFG on the undoped Cu(100) surface in relation to the temperature dependence of the "lattice" contribution to the EFG in the framework of the universal correlation found by Raghavan *et al*. for noncubic metals.

## II. METODOLOGY

To simulate the Cu(001) surface, we employ the slab-supercell (SC) approach, starting from the FCC bulk system, as shown in Fig. 1. We use a slab containing seven *inequivalent* substrate atomic layers, since this size (13 layers per SC) is sufficient to obtain a bulk-like environment at the deepest Cu atoms. In addition, we considered up to 11 *inequivalent* substrate layers to check size convergence. The thickness of the vacuum region between adjacent slabs was set to approximately 10 Å to prevent electronic coupling between the slabs. We call the *just-generated surface* the SC without any relaxation of its atomic positions; therefore, the forces on the Cu atoms are much stronger as they approach the surface. Consequently, we performed a complete relaxation of the atomic positions in the just-generated SC, yielding the equilibrium structure and the *reconstructed surface*. The notation $Cu_i$ (i = 1–7) refers to Cu atoms from the surface to the bulk, with $Cu_1$ being the topmost Cu atom and $Cu_7$ the deepest one. To account for thermal effects, we use experimentally determined lattice parameters for Cu bulk as a function of temperature over the 20-1350 K range [20]. Using these parameters, we begin by obtaining the reconstructed surface from the just-generated one at 20 K. At these structural equilibrium positions, we adjust the lattice parameters to those corresponding to the next higher temperature and perform a complete relaxation again, yielding a new reconstructed surface for that temperature. This procedure is repeated, increasing the temperature until the parameters corresponding to the highest temperature are used, as in an experiment in which the measurement temperature is increased gradually. For verification, we also relaxed the just-generated SCs obtained using the lattice parameters at each temperature. The $V_{33}$ values obtained in this way at the $Cu_1$ atom were very similar to those obtained with the previous procedure.

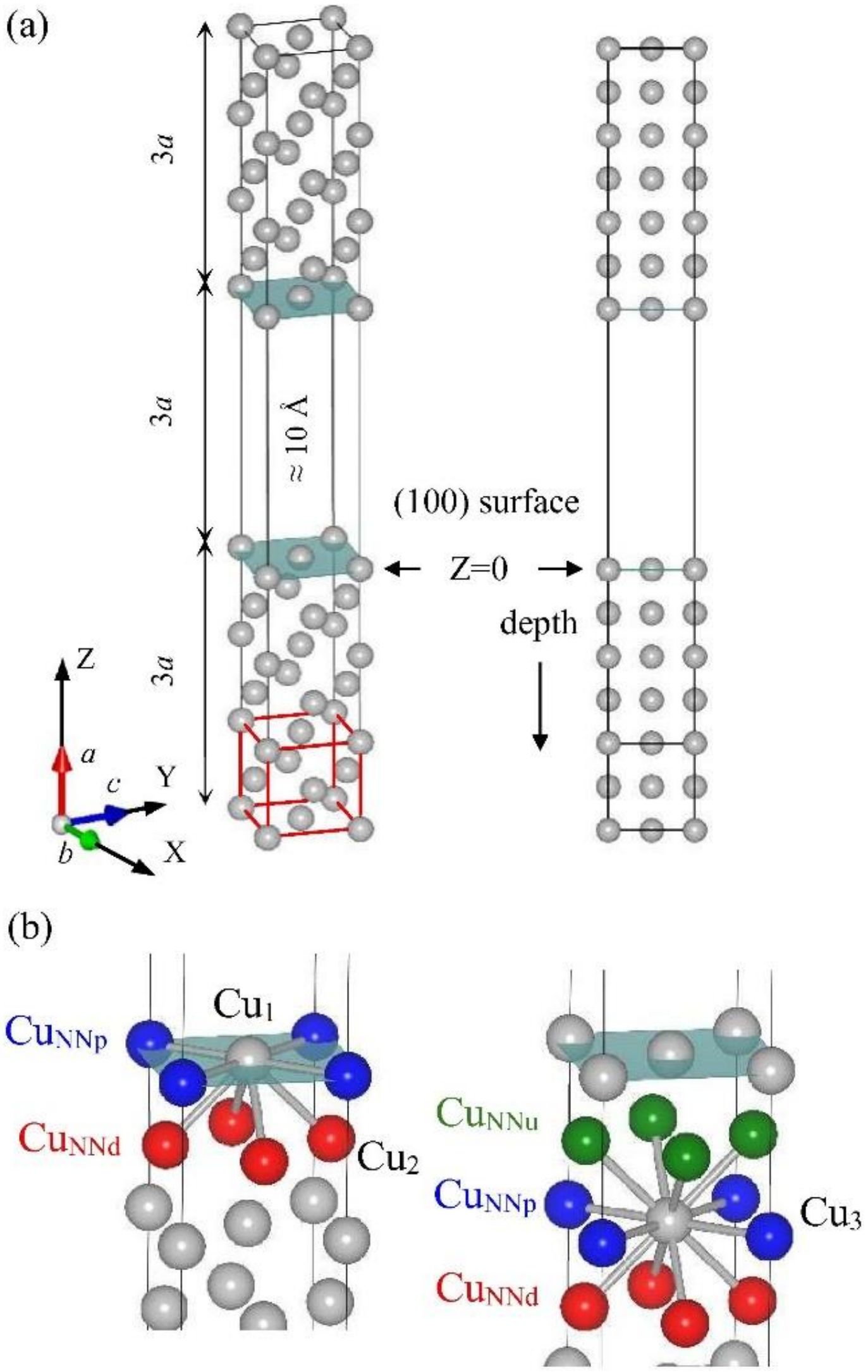

**Figure 1.** (a) Slab-supercell containing seven inequivalent substrate layers (13 atomic layers in total), used to simulate the Cu(100) surface. The FCC cell is shown in red. Crystalline axes and principal axis system of the diagonalized EFG are shown. (b) Zoom of the surface region. Left: the eight nearest-neighbor $Cu_{NN}$ atoms to Cu1 are divided into four $Cu_{NNp}$ in the same surface plane (blue atoms) and four $Cu_{NNd}$ in the lower-layer plane (red atoms). Right: the twelve $Cu_{NN}$ atoms to $Cu_3$ are divided into four $Cu_{NNp}$ in the same layer plane (blue atoms), four $Cu_{NNd}$ in the lower-layer plane (red atoms), and four $Cu_{NNu}$ in the upper-layer plane (green atoms).

In this study, we use state-of-the-art first-principles calculations within the framework of DFT [21], [22], employing the full-potential augmented plane-wave plus local orbitals (APW+lo) method [23], implemented in the WIEN2k package [24]. Exchange-correlation effects were treated using the Perdew-Burke-Ernzerhof (PBE-GGA) parametrization [25]. The cutoff parameter of the plane-wave basis is $R_{MT}K_{max} = 7$, where $R_{MT}$ and $K_{max}$ are the chosen radius of the nonoverlapping muffin-tin Cu spheres and the greatest modulus of the lattice vectors in reciprocal space, respectively. We set the radius of the sphere centered on the Cu atom to $R_{MT}$(Cu) = 1.17 Å. The tetrahedron method [26] was used to integrate in reciprocal space, with a *k*-space grid of 14x14x2. Finally, the equilibrium structures for all calculations were obtained using a Newton-damped scheme until the ion forces were below 0.01 eV/Å. The EFG tensors were computed from the second derivative of the full electric potential evaluated for the final equilibrium structures [27], [28].

## III. RESULTS AND DISCUSSION

In this section, we study the structural, electronic, and hyperfine properties of the *just-generated* and *reconstructed* Cu(100) surfaces to help elucidate the unexpected linear temperature dependence of the EFG for deposited ($^{111}$In→)$^{111}$Cd impurities on the Cu(100) surface [10]. In this first stage, we isolate the effects arising solely from the surface boundary conditions.

Due to these boundary conditions, only the topmost Cu atom ($Cu_1$) has eight Cu atoms as nearest neighbors: four ($Cu_{NNp}$) in the same surface plane and four ($Cu_{NNd}$) in the lower layer (see Fig. 1(b), left). The remaining $Cu_i$ atoms (i from 2 to 7) have twelve nearest neighbors: four in the same plane ($Cu_{NNp}$), four in the lower layer ($Cu_{NNd}$), and four in the upper layer ($Cu_{NNu}$), as in bulk FCC copper. This coordination is illustrated for the $Cu_3$ atom on the right-hand side of Fig. 1(b).

When the surface reaches its equilibrium structure, the system relaxes, and the distances between each inequivalent $Cu_i$ atom and its nearest neighbors change. In Fig. 2, we plot the distance d($Cu_i$-$Cu_{NN}$) between each inequivalent $Cu_i$ atom in the SC and its neighbors as a function of temperature for the *just-generated surface* (hollow squares) and the *reconstructed surface* after the converged full relaxation (triangles and circles).

As shown, for all inequivalent $Cu_i$ atoms, the distances to the $Cu_{NNp}$ atoms, which lie in the same plane as $Cu_i$, remain unchanged as the structure relaxes (at 0 K) to the *reconstructed* surface. That is, the atomic positions of the bulk system for these four $Cu_{NNp}$ atoms are preserved. This will not be the case for the coordinated atoms outside this plane during relaxation. Also, the d($Cu_i$-$Cu_{NNp}$) distances (and angles) exhibit the same isotropic temperature dependence as in the bulk (blue circles and hollow squares are superimposed at all temperatures).

The $Cu_1$ atom is the only one without nearest-neighbor atoms above it. The d($Cu_1$-$Cu_{NNd}$) distance is lower on the *reconstructed* surface than on the *just-generated* one at all temperatures. For temperatures greater than 300 K, this distance is nearly constant, in contrast to the behavior of d($Cu_1$-$Cu_{NNp}$). In fact, a small accordion effect is superimposed on the nearly constant temperature dependence of this bond length. This effect should arise from the interplay between the (nonlinear) lattice-parameter expansion and the linear contraction of the distance between the surface and the 2nd atomic layer, as we explain below.

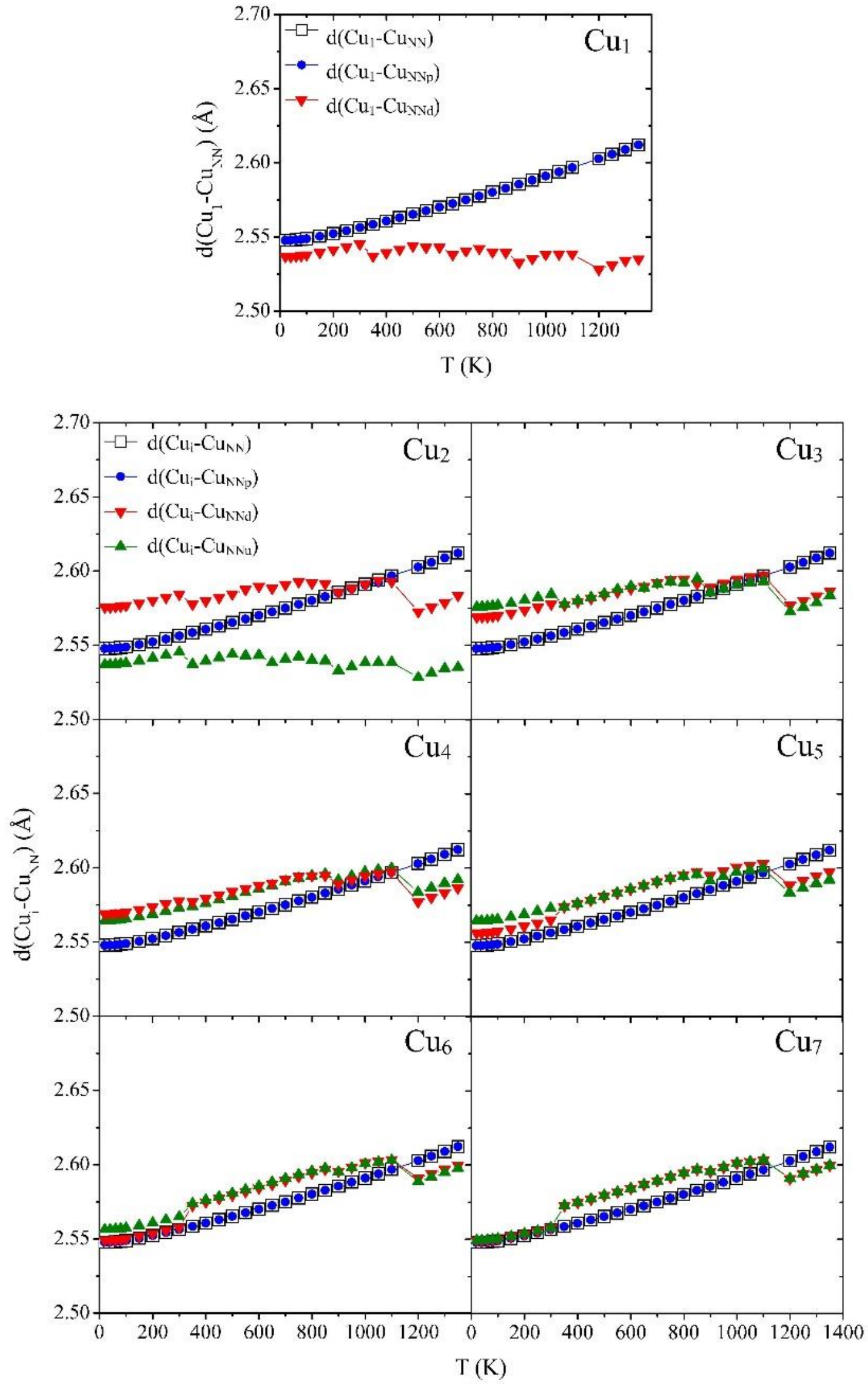

**Figure 2.** Distance d($Cu_i$-$Cu_{NN}$) for each inequivalent $Cu_i$ atom in the SC for the *just-generated* surface (hollow squares) and the *reconstructed* surface (triangles and circles), as a function of temperature (up: i=1, and down: i from 2 to 7). The blue circles and red and green triangles correspond to $Cu_{NN}$ on the same plane and on the lower- and upper-layer planes relative to the $Cu_i$ plane, respectively.

$Cu_2$ is the atom in the layer immediately beneath the surface. For this atom, the behavior of d($Cu_2$-$Cu_{NNu}$) and d($Cu_2$-$Cu_{NNd}$) as temperature increases is essentially the same, with the difference that d($Cu_2$-$Cu_{NNu}$) is smaller than d($Cu_2$-$Cu_{NNp}$), whereas d($Cu_2$-$Cu_{NNd}$) is greater than, equal to, and smaller than d($Cu_2$-$Cu_{NNp}$) from 20 to 900 K, from 900 to 1100 K, and from 1100 to 1350 K, respectively. As the $Cu_i$ atom "moves" away from the surface, i.e., from i=3 to i=7, d(Cui-CuNNu) and d(Cui-CuNNd) approach d(Cui-CuNNp). For $Cu_7$, they become equal (as in Cu bulk) from 20 to 300 K and are practically the same (difference less than $10^{-2}$ Å) at temperatures above 300 K, indicating reconstruction of the bulk environment.

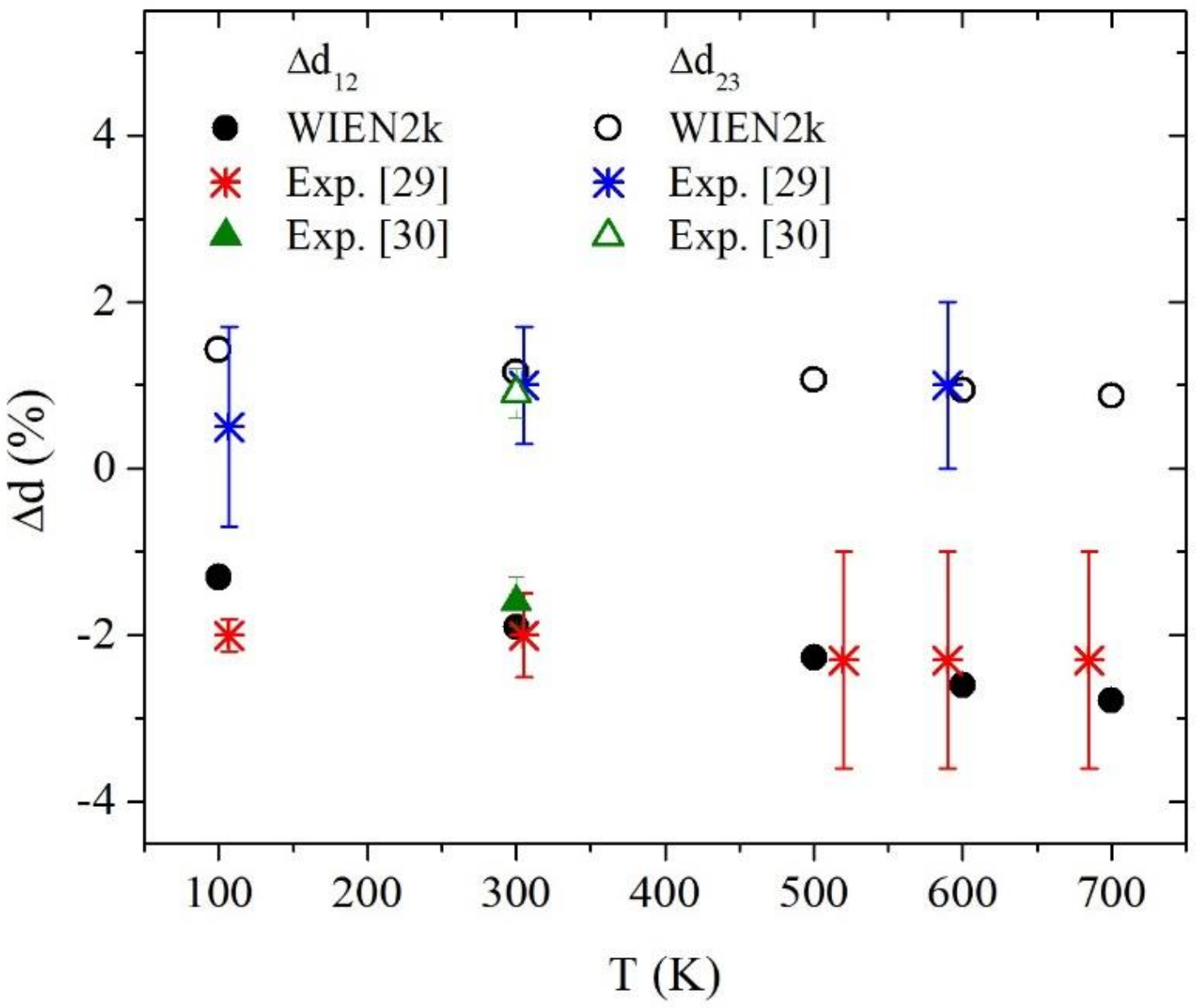

**Figure 3.** Predicted relative displacement Δd of the topmost ($\Delta d_{12}$) and second ($\Delta d_{23}$) interlayer distances of the *reconstructed* surface with respect to the interlayer distance in Cu bulk at each temperature. The experimental data come from Refs. [29] and [30].

Figure 3 plots the relative variation Δd of the topmost ($\Delta d_{12}$) and second ($\Delta d_{23}$) interlayer distances as a function of temperature for the *reconstructed* surface, with respect to the interlayer distance in Cu bulk at each temperature. A positive (negative) value of Δd indicates expansion (contraction) of the interlayer distance upon relaxation at each higher?? temperature. In general, the topmost 1-2 interlayer distance contracts, while the second 2-3 interlayer distance expands relative to the bulk value, in excellent agreement with experimental determinations [29],[30], pointing to a good description of the *reconstructed* surface. The bond length d($Cu_1$-$Cu_{NNd}$), smaller than the bulk values (Fig. 2, up), is consistent with the contraction of the 1-2 interlayer of the *reconstructed surface*, already at 0 K. This bond length constant dependency on temperature (see Fig. 2) indicates that, while the 1-2 interlayer distance decreases, the angles of these bonds relative to the surface plane also decrease. Contrarily, the expansion of the 2-3 interlayer enables the expansion of the bond length d($Cu_2$-$Cu_{NNd}$) at low temperatures relative to the bulk value and also remains rather constant with temperature [presenting a similar accordion effect as in d($Cu_2$-$Cu_{NNu}$)]. The 2-3 initially expanded interlayer spacing decreases slightly as temperature increases.

We now present the EFG predictions at a representative temperature. Figure 4 shows the $V_{33}$ values (the largest component of the diagonalized EFG tensor) for $Cu_i$ atoms at different depths (Z) and as a function of the number *i* of inequivalent atomic slab layers used in the *reconstructed* Cu(100) surface, using Cu lattice parameters at 300 K. As shown, $Cu_7$ atoms located seven slab layers below the Cu(100) surface (Z ≈ 11 Å) exhibit a null $V_{33}$ value, recovering the bulk FCC symmetry of their environment and no longer experiencing the surface. For all EFG calculations at each $Cu_i$ atom, we obtained $\eta_i$=0 (axial symmetry), and $V_{33}$ is aligned with the [100] direction (i.e., perpendicular to the surface). Consequently, all remaining *ab initio* results presented here were obtained using a slab with seven inequivalent atomic layers.

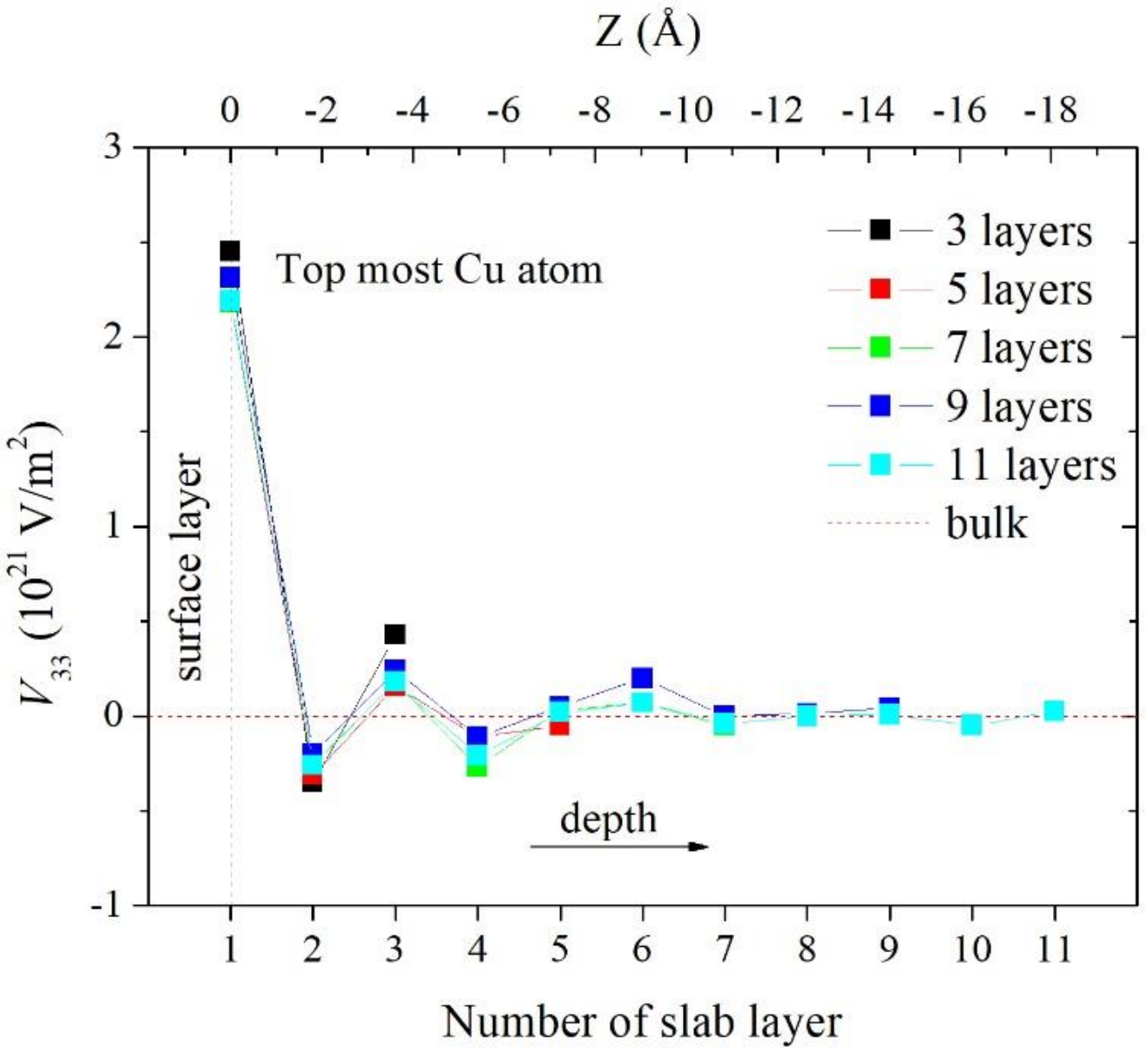

**Figure 4.** Predicted V33 value as a function of the depth of the Cu atom from the reconstructed Cu(100) surface, for different numbers of inequivalent atomic slab layers used in the calculations at 300 K (from 3 to 11 inequivalent layers, indicated with different colors). The dashed red line represents the V33 value for bulk Cu. Z = 0 Å is the position of the topmost Cu atom in the reconstructed Cu(100) surface.

As shown in Fig. 4, there are essentially two values for $V_{33}$, one ($V_{33}^{Surface}$) for the topmost $Cu_1$ atom and another ($V_{33}^{B}$) for the remaining $Cu_i$ atoms (from i=2 to 7) located below the surface. Comparing $V_{33}$ at Cu atoms in the 2$^{nd}$ and 3$^{rd}$ layers with that of the topmost $Cu_1$ atom at the surface yields a factor of approximately 2. We cannot compare $V_{33}^{Surface}$ to the bulk value because it is null; however, a significant increase in $V_{33}$, approximately a factor of 5, was experimentally observed at $^{111}$Cd probes on the In(111) metallic surface relative to its bulk value [31]. A thorough inspection of the electron densities shown in Fig. 5 is representative and visually explains the increase in $V_{33}$ from the bulk to the Cu(100) surface. This figure shows the electron density $\rho(\vec{r})$ at the (010) and (100) planes containing the topmost $Cu_1$ atom and the deepest $Cu_7$ atom in the *reconstructed* surface, for the most energetic (conduction) electron, calculated at 300 K. The critical angle $\theta_c$ = 54.7° (Fig. 5, up), measured from the $V_{33}$ axis, separates regions of negative charge that contribute to $V_{33}$ with opposite signs (see Eq. (14) of Ref. [32] for details): negative charge localized below $\theta_c$ (in fact, a conical region in three dimensions with $V_{33}$ as its symmetry axis) contributes to $V_{33}$ with negative values, and vice versa. At $Cu_7$, the four elongated spots contribute to $V_{33}$ with both positive and negative signs, canceling each other. At the surface, the spatial distribution of this negative charge at $Cu_1$ changes strongly, with almost all the charge distributed with axial symmetry around the $V_{33}$ axis and located in the (100) plane ($\theta \approx 90°$), thus contributing to a *large positive* EFG, in contrast to the almost null value at the $Cu_7$ atom.

We chose to first discuss the origin of the EFG at the surface using the last electron for several reasons. In the EFG parametrization for noncubic metals proposed by Raghavan *et al.* [33], [34], [7], the conduction-electron contribution to $V_{33}$ has about two to three times the magnitude of the rest of the EFG sources. In effect, this parametrization proposes:

$$V_{33}^{tot} = V_{33}^{ionic} + V_{33}^{elect} = (1-\gamma_\infty)V_{33}^{latt} + V_{33}^{elect} \,. \qquad (1)$$

In Eq. 1, the *ionic* contribution comes from the external EFG produced at the probe nucleus by the rest of the atoms of the crystal lattice, $V_{33}^{latt}$, plus another term arising from the antishielding effect of $V_{33}^{latt}$ produced by the probe´s filled electronic core shells, $-\gamma_\infty V_{33}^{latt}$, being $\gamma_\infty$ the Sternheimer antishielding factor [35]-[37]. $V_{33}^{elect}$ is the *electronic* contribution arising from conduction electrons. If the empirical "universal correlation" [33], [34], [7] found for many noncubic metals applies for Cu and its surface, $V_{33}^{elect}$ is proportional, and with opposite sign, to $V_{33}^{ionic}$ :

$$V_{33}^{elect} = -K(1-\gamma_\infty)V_{33}^{latt} \;, \tag{2}$$

being $K$ and $\gamma_\infty$ positive and negative numbers, respectively. In this case $V_{33}^{tot}$ reduces to:

$$V_{33}^{tot} = (1-K)(1-\gamma_\infty)V_{33}^{latt} = \; A_{Cu}\; V_{33}^{latt} \approx -2(1-\gamma_\infty)V_{33}^{latt} \,. \tag{3}$$

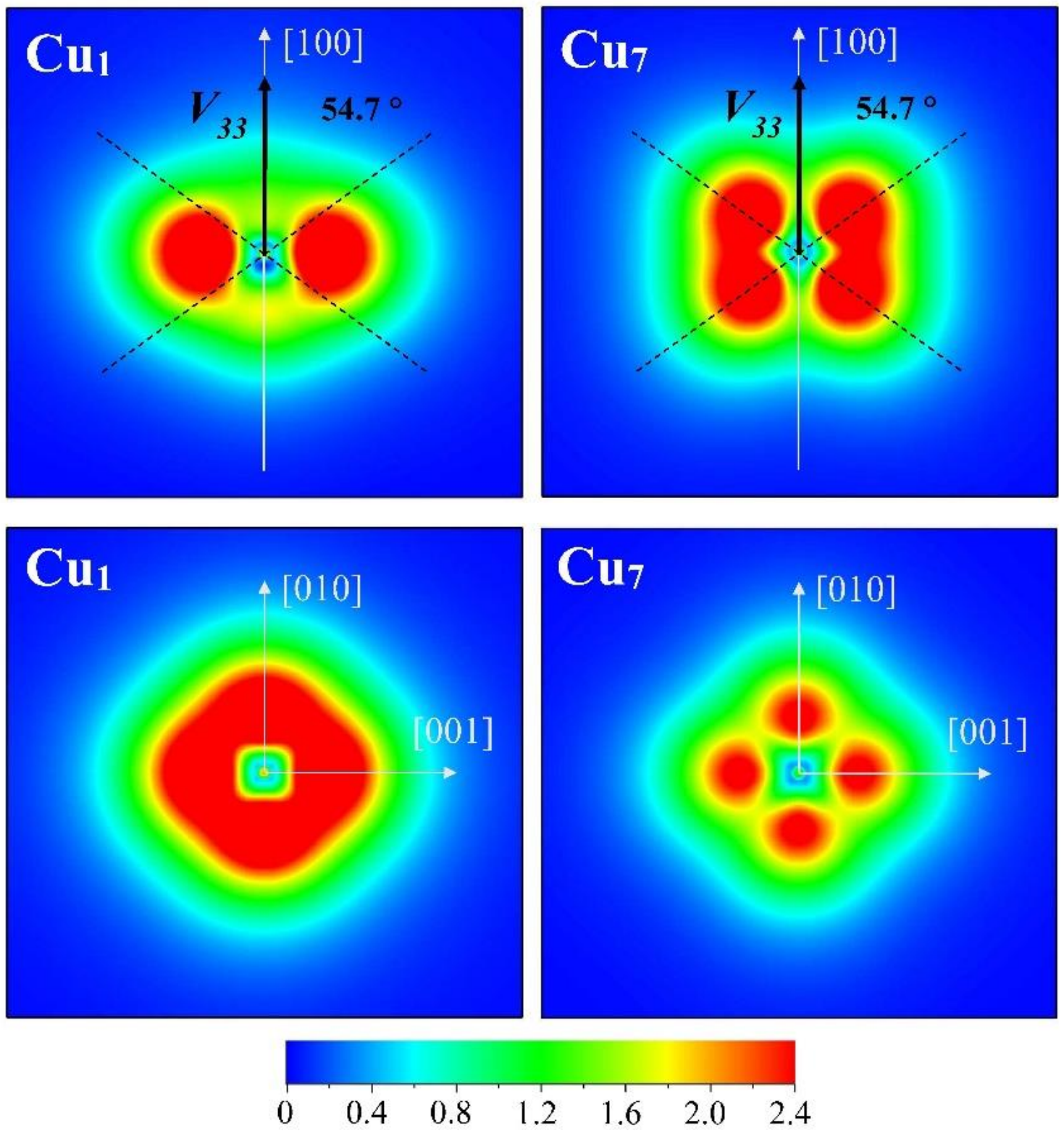

**Figure 5.** Electron density $\rho(\vec{r})$ at the (010) (up) and (100) (bottom) planes containing the topmost $Cu_1$ atom and the deepest $Cu_7$ atom of the *reconstructed* Cu(100) surface, for the most energetic (conduction) electron at 300 K.

In Eq. (3), we have used the fact that for many metallic impurity-host systems $K \approx +3$, in particular for small and medium values of $V_{33}^{ionic}$, but other metals showed $K \approx +2$. Hence, we can suppose that the conduction-electron contribution to the EFG may be dominant and representative of the total EFG. As we will discuss later, $V_{33}$ "lattice" calculations at the surface ($Cu_1$ site) yield a negative value, hence the reversal sign of the universal correlation applies to this surface system (since, as Fig. 4 shows, the EFG at the surface is positive), with a multiplicative coefficient $A_{Cu} = (1-K)(1-\gamma_\infty)$ around twice or equal (depending on $K$) to that predicted by the Sternheimer antishielding factor [35]-[37]. On the other hand, the electron densities in Fig. 5 were constructed by filtering the partial density of states (PDOS) projected onto the $Cu_1$ atom over an energy range 1.47 eV below the Fermi level, corresponding, as mentioned before, to the more energetic (conduction) electron.

In Fig. 6, we show the total DOS for the *reconstructed* Cu(100) surface at 300 K, the *s*-, *p*-, and *d*-PDOS projected onto $Cu_1$, and a zoom-in on the $p_z$ and $p_x+p_y$ contributions at the topmost atom. Successive energy regions corresponding to the last, 2nd, 3rd, and 4th more energetic electrons are separated by vertical lines. The $Cu_1$-$p$ and $Cu_1$-$d$ contributions to the total EFG at the surface are positive and negative, respectively, with the $p$ contribution dominant. The majority contribution to the EFG can be split in two parts [18], [27], [28], [38], $V_{33}^{p}$ and $V_{33}^{d}$, proportional to the $p$ and $d$ asymmetry counts $\Delta n_p$ and $\Delta n_d$, which are given in terms of the partial charges (occupation numbers) $n_{p_i}$ and $n_{d_i}$ of the corresponding $l_i$ orbital:

$$\Delta n_p = \frac{1}{2}\left(n_{p_x} + n_{p_y}\right) - n_{p_z} \tag{4}$$

$$\Delta n_d = \left[\left(n_{d_{x^2-y^2}} + n_{d_{xy}}\right)\right] - \left[\frac{1}{2}\left(n_{d_{xz}} + n_{d_{yz}}\right) + n_{d_{z^2}}\right] \quad . \tag{5}$$

As seen in Fig. 6 (bottom), the energy region of the last electron reflects the positive character of the $p$ contribution to $V_{33}$ that holds across all energies, with the $d$ contribution also (locally) positive in this small energy region. On the other hand, the magnitude of $V_{33}$ depends mainly on the distance of the nonspherical source charges to the probe nucleus. In the Cu atom, the first node of the $3p$ wave function (and even more if $4p$ states are also populated in the solid) is at shorter distance than the one of the $3d$ wave function, according to the rule that the number of nodes is equal to n-l-1, resulting in a dominant $p$ contribution to $V_{33}$ even though the asymmetry count $\Delta n_d$ may be larger than $\Delta n_p$. All these facts correlate perfectly with the angular shape of $\rho(\vec{r})$ shown in Fig. 5 at $Cu_1$ and justifies the choice of this energy region to illustrate the origin of the *large* and *positive* EFG at the surface. On the contrary, for Cu atoms in bulk, all shells must be completely spherical due to the crystal symmetry (all occupation numbers $n_{p_i}$ and $n_{d_i}$ are equal to 2 in Eqs. 4 and 5), leading to a null $V_{33}$ value.

Figure 7 shows plots of $\rho(\vec{r})$ for the 2nd-, 3rd-, and 4th-most energetic electrons at the topmost $Cu_1$. This sequence shows a transition from the electronic donut-like shape shown in Fig. 5 (left) at $Cu_1$ to geometries more and more aligned to the $V_{33}$ axis, contributing with a negative sign to the EFG in these energy regions, in accordance with the large and increasing number of $d$-states in the PDOS as going to lower energies (as Fig. 6 bottom showed). For the 4$^{th}$ electron, a small density perpendicular to $V_{33}$ is detected (Fig. 7, right), probably correlated with an increase in the occupation of the $p_x$ and $p_y$ states relative to the $p_z$ states. Nevertheless, the close proximity of the asymmetric $p$-states to the nucleus and the near parity of the total $d$ and $p$ asymmetry counts across all energies ensure a *positive* EFG at the Cu(100) surface.

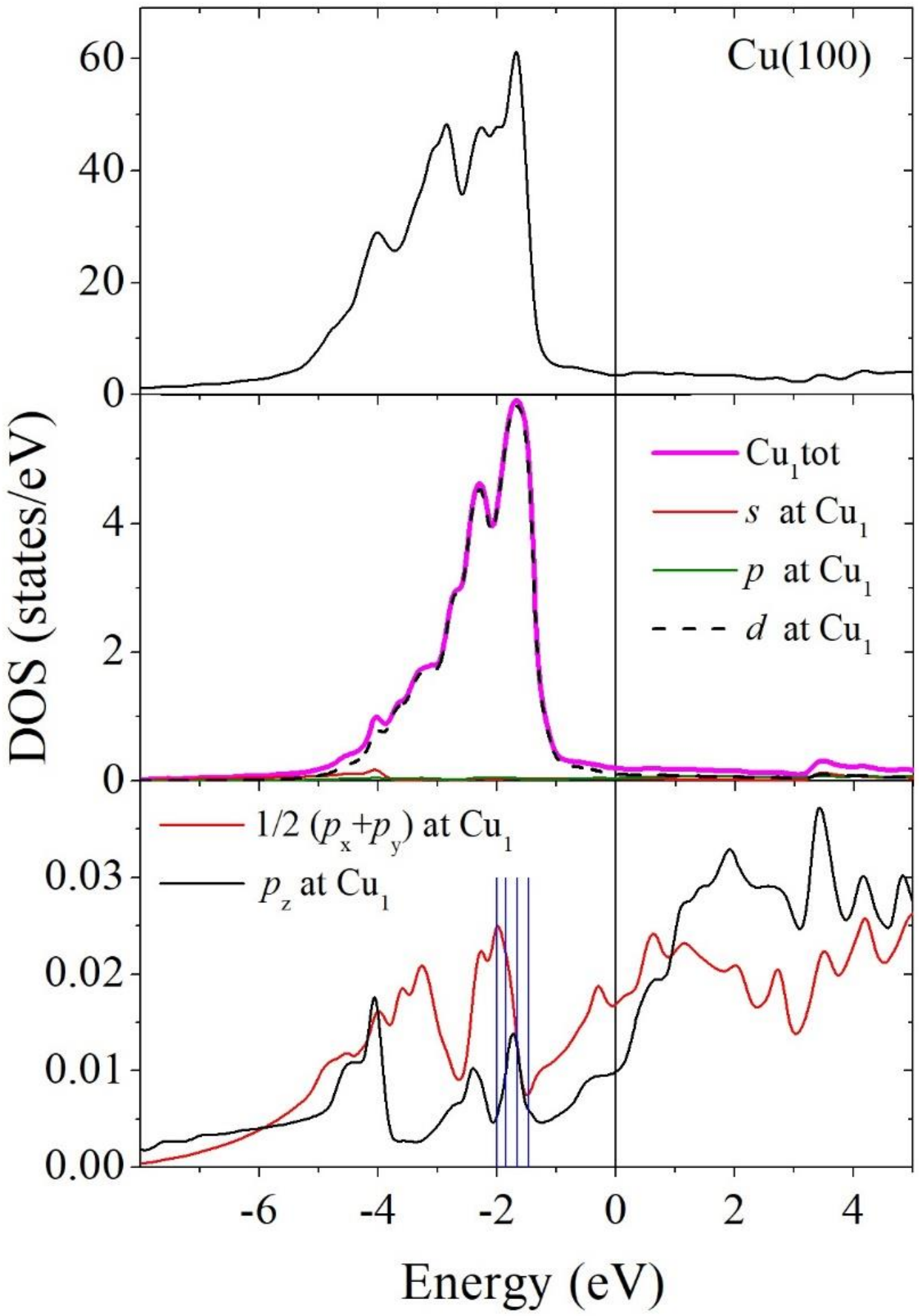

**Figure 6.** Electronic density of states (DOS) for the reconstructed Cu(100) surface at 300 K. Top: total DOS. Middle: s-, p-, and d-partial DOS projected onto Cu1. The p contribution is nearly invisible on this scale. Bottom: zoom on the pz and px+py contributions at this topmost atom. Energy is referenced to the Fermi level.

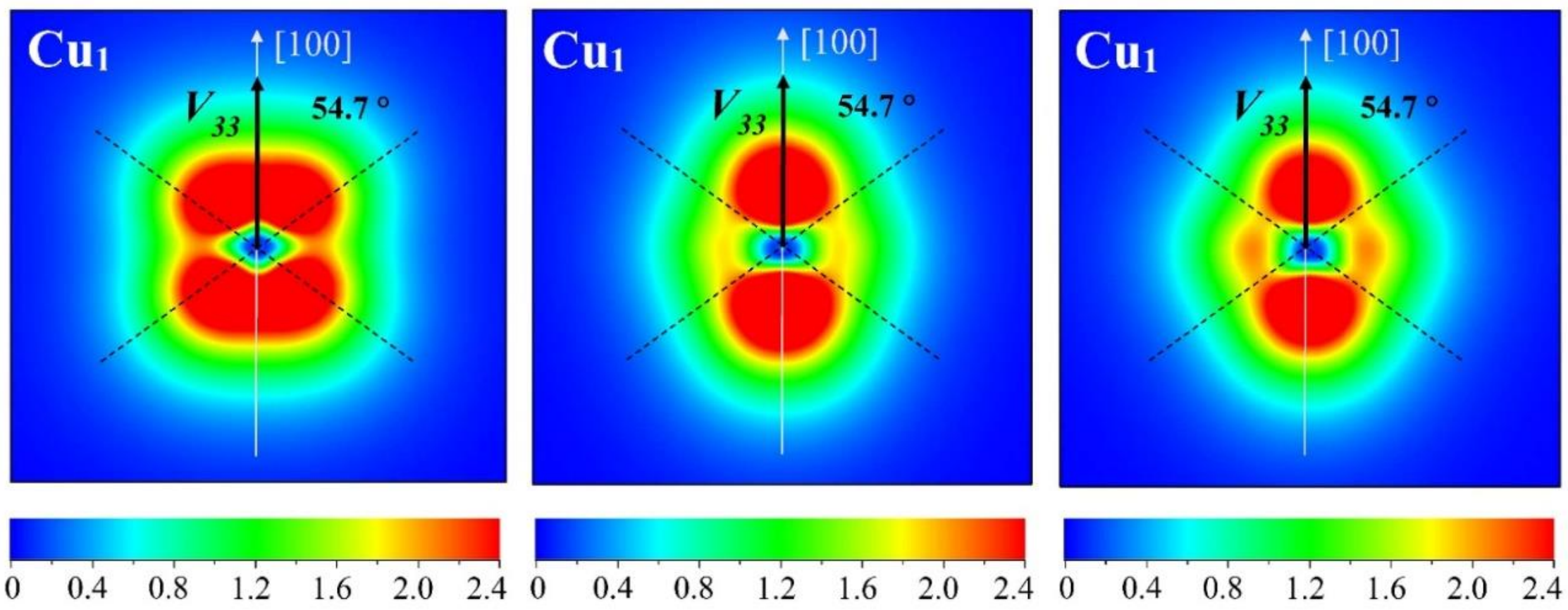

**Figure 7.** Electron density $\rho(\vec{r})$ in the (010) plane containing the topmost $Cu_1$ atom of the *reconstructed* Cu(100) surface, for the 2nd-most energetic electron (left), the 3rd- most energetic electron (middle), and the 4th-most energetic electron (right) at 300 K.

Now, we present the predictions of the EFG at the topmost $Cu_1$ atom ($V_{33}^{Surface}$) as a function of temperature. In Fig. 8, we plot our *ab initio* predictions for $V_{33}^{Surface}$ for both the *just-generated* and *reconstructed* Cu(100) surfaces (filled symbols). In both cases (for the just-generated and the reconstructed surfaces), $V_{33}$ varies almost linearly with temperature, as the experimental EFG at $^{111}$Cd probes at the Cu(100) surface does. When the (100) *reconstructed* surfaces reach their equilibrium structures, the predicted $V_{33}$ slope decreases slightly, and $V_{33}$(0 K) is larger than that of the *just-generated* surface. Both effects can be explained in terms of what happens with the distances $Cu_1$-$Cu_{NNd}$ (see Fig. 1.b and 2) and the angle $Cu_{NNd}$ forms with the [100] axis when the structure relaxes at each temperature. In effect, after relaxation at 0 K, the $Cu_1$-$Cu_{NNd}$ distance becomes shorter, but the angle increases, which are opposite behaviors for the EFG (Eq. (14) in Ref. [32]). In this case, the angle effect is stronger, reducing the negative NNd contribution to $V_{33}$, thus producing a larger total positive EFG for the *reconstructed* surface at 0 K. At higher temperatures, this effect becomes stronger. This is a direct consequence of the contraction of the interlayer distance $d_{12}$ as the just-generated surface relaxes, as shown in Fig. 3. These results support the conclusion that, at first order, the linear temperature dependence of $V_{33}$ at $^{111}$Cd atoms deposited onto Cu(100) surfaces arises from the surface generation and its reconstruction.

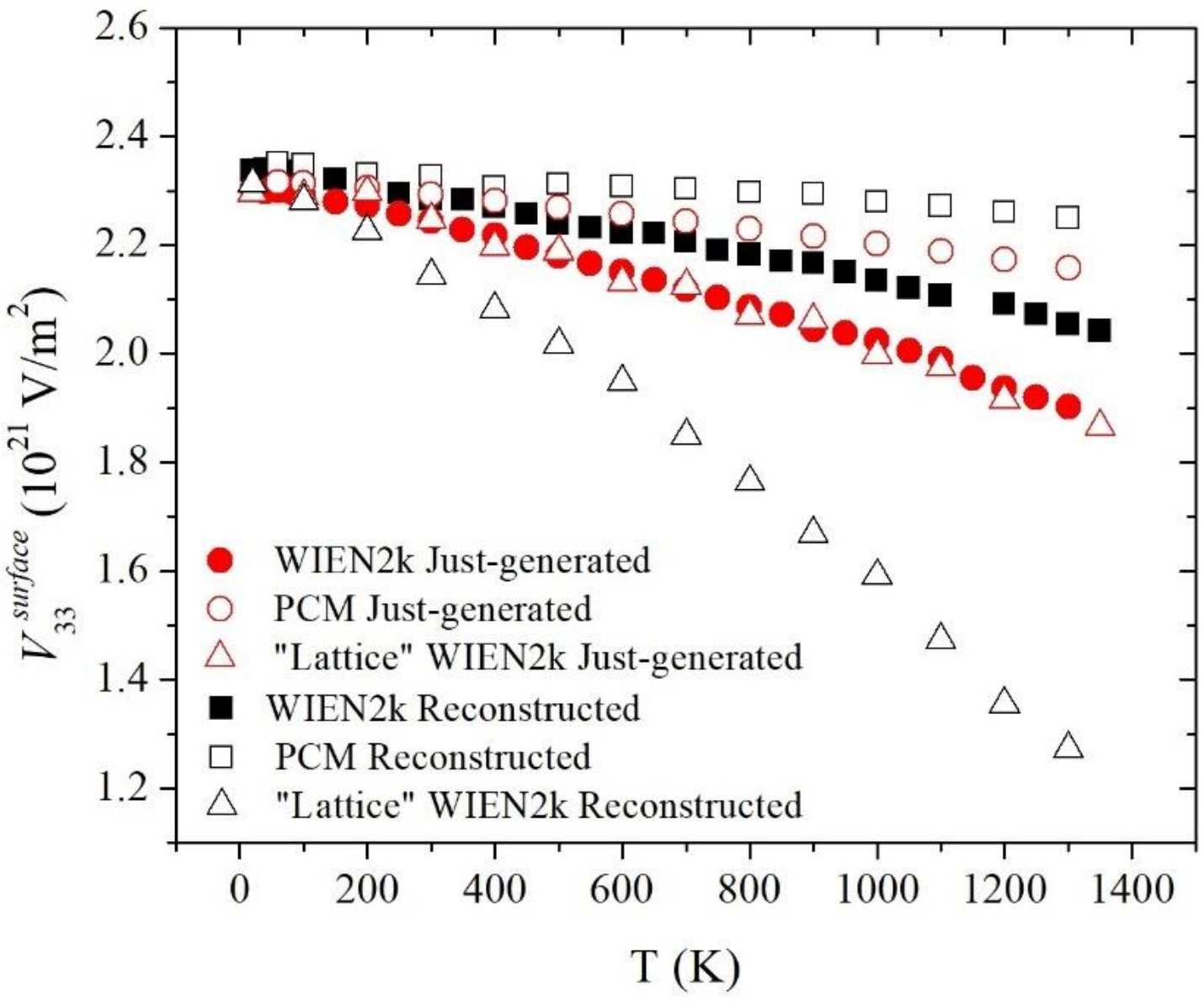

**Figure 8.** *Ab initio* predicted $V_{33}^{Surface}$ at the topmost $Cu_1$ atom for the *just-generated* and *reconstructed* Cu(100) surfaces, using Cu lattice parameters as a function of temperature (filled symbols). Open symbols are predictions of $V_{33}^{Surface}$ proposing a factorization as in Eq. (3): $V_{33}^{Surface}$ (T) = $A_{Cu}$ . $V_{33}^{latt}$ $(T)$, with the lattice contribution calculated using a point-charge model (PCM), considering first- and second-nearest neighbors from $Cu_1$, or the *interstitial* contribution to $V_{33}$ as computed by WIEN2k.

In Fig. 8, the predictions of different "lattice" contributions to $V_{33}$ are also shown using both the *just-generated* and *reconstructed* structures for analyzing the predicted $V_{33}$ behavior for the pure surfaces in the framework of Eq. 3. The hollow squares and circles are the results of calculating $V_{33}^{latt}$ using a simple point-charge model (PCM) taking into account the first and second nearest neighbors to the $Cu_I$ atom for both surfaces, using charge state q=1+ for all Cu ions in the lattice. Calculations including next-nearest neighbors give similar results, showing convergence within the PCM. The results shown here were obtained using, in Eq. 3, a multiplicative constant $A_{Cu} \approx -10$, hence obtaining a $K$ parameter closer to 2 in this approach (being the nominal value $\gamma_{\infty}$= 17.37 for Cu [37]). If the "lattice" contribution is instead represented by the "interstitial" contribution to $V_{33}$ calculated using WIEN2k, which accounts for the EFG contribution coming from the other Cu´s muffin-tin spheres and the entire interstitial region, the match is perfect when the *just-generated* structure is used. In this case, $A_{Cu} \approx -40$, giving a K parameter closer to 3. Conversely, if the *reconstructed* structure is used, the slope obtained is too strong.

Nevertheless, the qualitative agreement shown in Fig. 8 indicates that the linear dependence of $V_{33}^{Surface}$ on T in the pure Cu(100) surface is mainly due to the structural lattice behavior with T, already in the case of the *just-generated* surface. Also, it is apparent that the subsequent lattice relaxation decreases the slope of the linear T dependence, as we have discussed.

Now, fitting both *ab initio* predicted curves with $V_{33}(T) = B + M \cdot T = [V_{33}(0\ K) + M \cdot T] \times 10^{21}$ V/m$^2$, we found for the *just-generated* surface $V_{33}(0\ K) = 2.32$ and M = -3.1 x $10^{-4}$ (M/B= 0.013%), and for the *reconstructed* one $V_{33}(0\ K) = 2.35$ and M = -2.21 x $10^{-4}$ (M/B=0.0094%). For this temperature range, a linear fit to the experimental $V_{33}$ values reported by Klass *et al.*

[10], using a quadrupole moment Q = + 0.76(2) b [39], gives: $V_{33}$(0 K) = 11.25(5) and M = -11.8(5) x $10^{-4}$ (M/B= 0.0104(5) %). This surprising but apparent coincidence in the M/B ratio in three very different systems is also shown through the matching slopes of the normalized T dependencies displayed in Fig. 9 for these different structures. In this figure, the blue line is a least-squares fit to the experimental $V_{33}$ values reported in Ref. [10] for $^{111}$Cd atoms deposited onto the Cu(100) surface. Always within Raghavan´s parametrization of Eq. 3, and considering the linear dependence of the form b + m .T (b negative and m positive) found for $V_{33}^{lattice}$ for the *just-generated* and *reconstructed* surfaces, the physical meaning of M/B results to be the ratio m/b, i.e., the normalized slope of the *lattice* contribution to the EFG, which reflects a similar structural lattice behavior with T for the three systems.

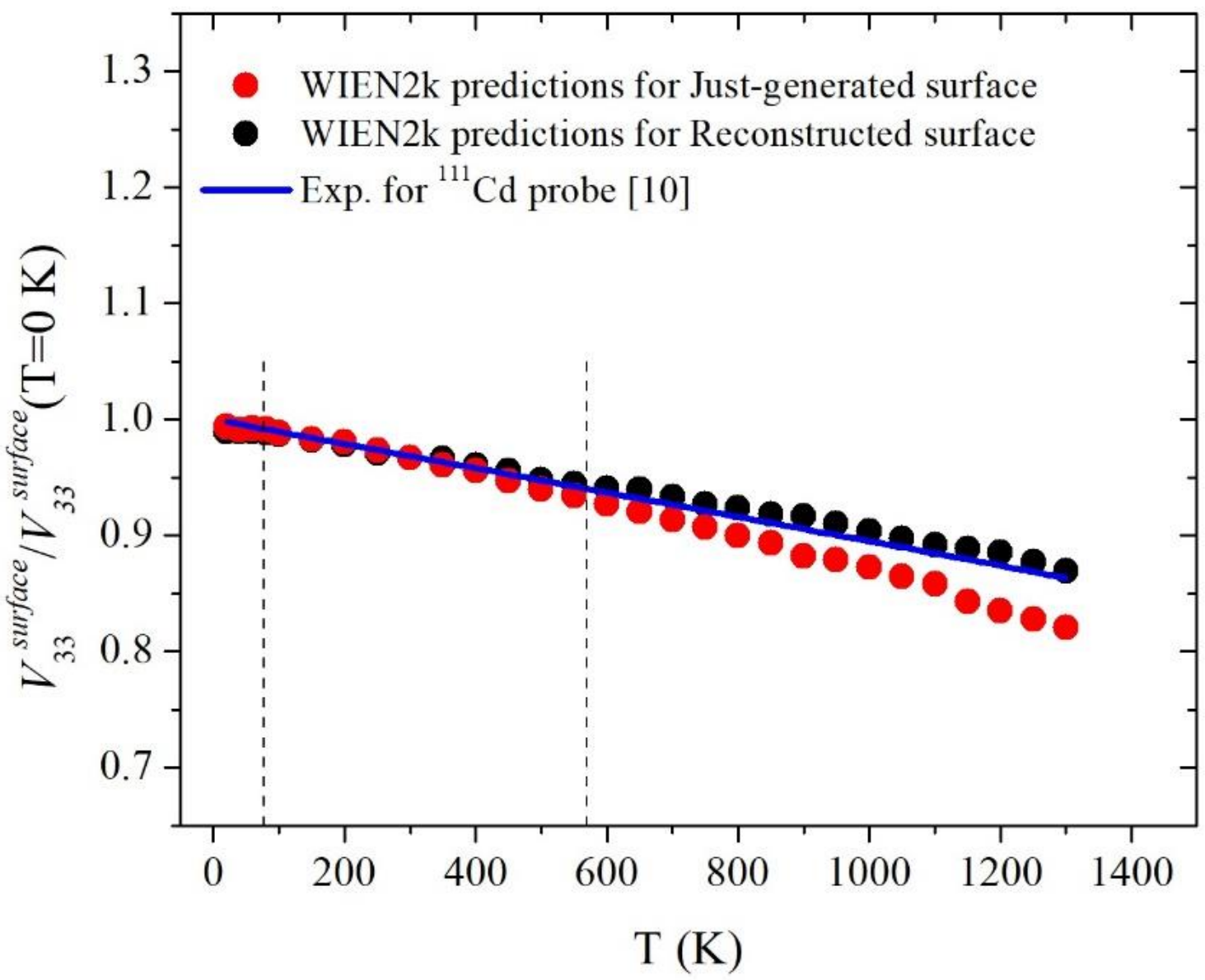

**Figure 9:** Normalized $V_{33}^{Surface}$ (at the topmost Cu1 atom) for the *just-generated* and *reconstructed* surfaces compared with a linear fit to the experimental $V_{33}$ values measured at ($^{111}$In→)$^{111}$Cd probes deposited onto a Cu(100) surface [10]. Vertical dashed lines denote the experimental T region.

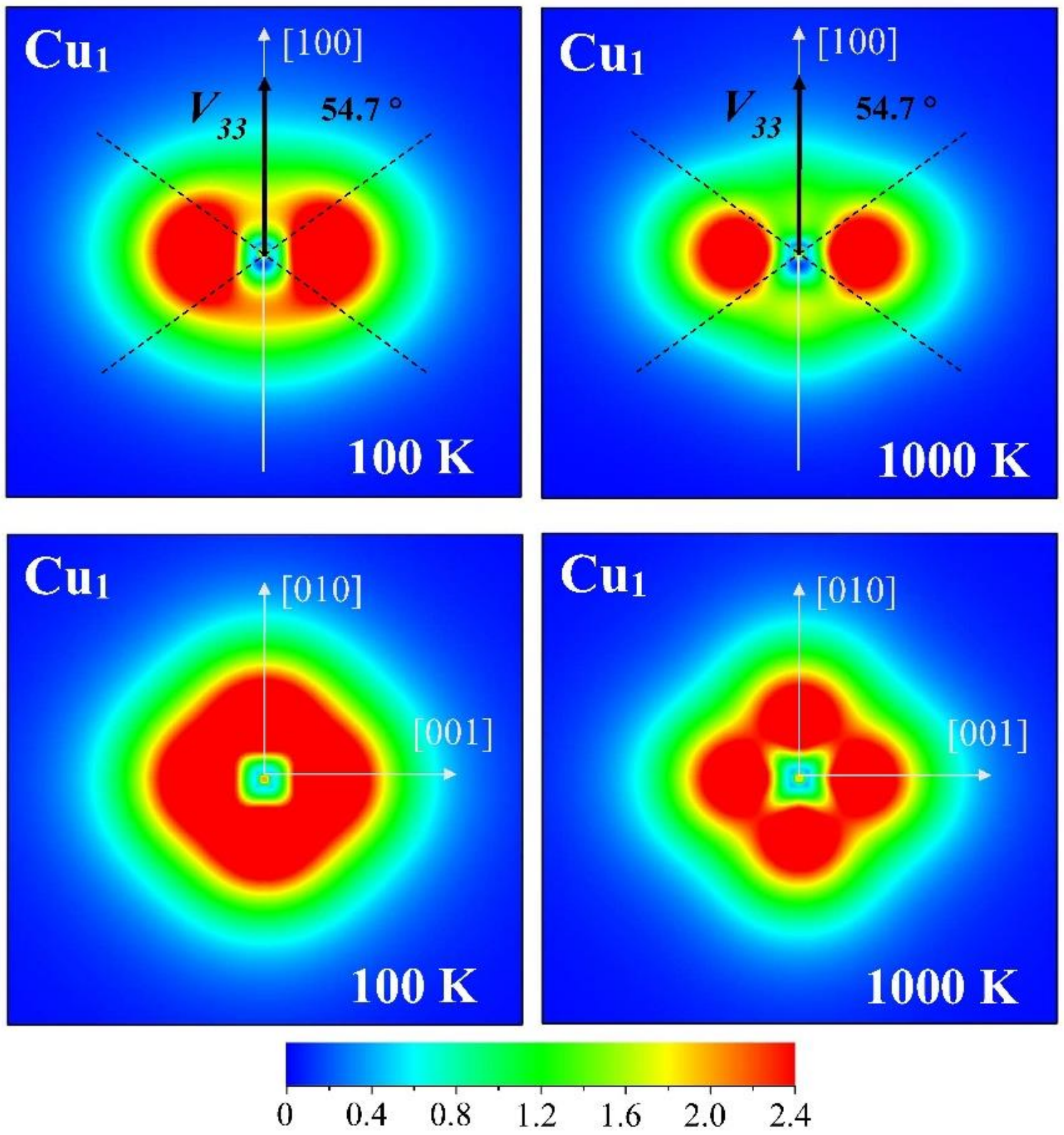

**Figure 10.** Electron density $\rho(\vec{r})$ at the (010) (up) and (100) (bottom) planes containing the topmost $Cu_1$ atom, for the most energetic (conduction) electron at 100 K and 1000 K.

Finally, Fig. 10 shows electron densities for the topmost $Cu_1$ atom, corresponding to the most energetic electron in the *reconstructed* surface, for extremely different temperatures. These densities yield positive $V_{33}$ values along the [100] axis at both temperatures, because the negative charge is mainly distributed closer to the (100) plane than to the $V_{33}$ axis, as explained earlier. In addition, the magnitude of this charge's contribution to $V_{33}$ decreases with increasing temperature. In effect, at 1000 K, the negative charge remains near the (100) plane but is substantially reduced relative to lower temperatures, in total agreement with the (linear) decrease of the EFG at the topmost $Cu_1$ atom of the (100) surface.

## IV. CONCLUSIONS

Applying state-of-the-art first-principles calculations, we studied the Cu(100) surface reconstruction and its structural, electronic, and hyperfine properties as a function of temperature and depth from the surface.

Using experimentally determined temperature-dependent lattice parameters of copper, we studied the reconstructed surfaces, finding that the contraction of the topmost 1-2 interlayer distance and the expansion of the second 2-3 interlayer distance relative to the bulk interlayer width are in excellent agreement with experimental results as a function of temperature, supporting the predicted reconstructed surfaces.

We found that the spatial redistribution of the electronic density of conduction electrons at the atomic scale near the nucleus of the topmost $Cu_1$ atom explains the surface effect on the large EFG value relative to the bulk value, and also its decrease with increasing temperature.

We demonstrated that $V_{33}$ at the topmost Cu atom in the *ab initio*-predicted *just-generated* and *reconstructed* surfaces, as well as in $^{111}$Cd atoms deposited onto the Cu(100) surface, all show a linear decrease with increasing temperature. We also show that the normalized slopes of these linear dependencies almost match between them, indicating that, within Raghavan´s EFG parametrization for metals, this agreement reflects similar lattice EFG behavior. As a consequence, we show that the linear decrease of the EFG as T increases on the $^{111}$Cd-doped Cu(100) surface, evidenced by the TDPAC experiments, is probably mostly caused by the generation of the pure Cu(100) surface and its subsequent reconstruction.

## AUTHOR INFORMATION

### Corresponding Authors

*E-mail: darriba@fisica.unlp.edu.ar ; renteria@fisica.unlp.edu.ar

## ACKNOWLEDGEMENTS

CONICET and UNLP partially supported this work under Research Grants No. PIP0901 and No. 11/X1004, respectively. This research used the computational facilities of the Physics of Impurities in Condensed Matter group at IFLP and the Department of Physics (UNLP). G.N.D. and M.R. are members of CONICET, Argentina. R.F. acknowledges PEDECIBA, CSIC-UdelaR, and ANII, all Uruguayan Institutions for financial support.

**TOC**

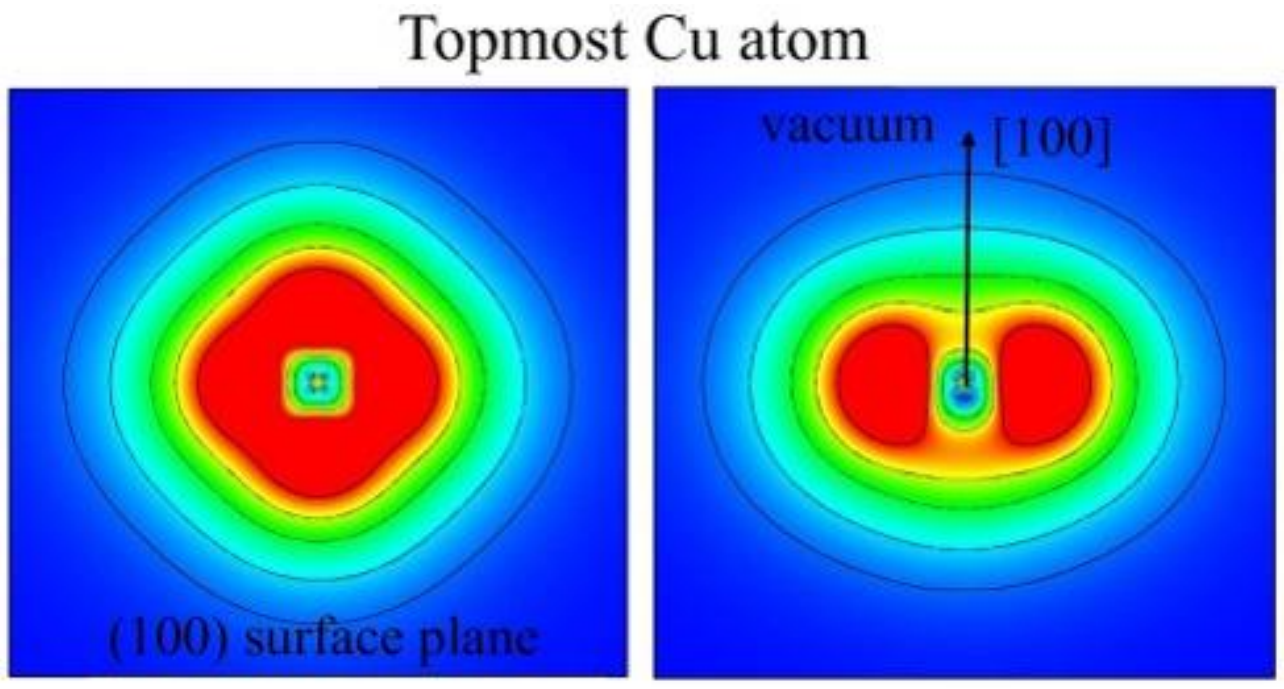